\documentclass[preprint,showpacs,preprintnumbers,amsmath,amssymb]{revtex4}

\usepackage{epsfig}
\usepackage{graphics}
\usepackage{subfigure}

\begin{document}

\def\Huge{\huge}
\def\e{\begin{equation}}
\def\f{\end{equation}}
\def\*{^{\displaystyle*}}
\def\=#1{\overline{\overline #1}}
\def\d{\partial}
\def\s{\strut\displaystyle}
\def\-#1{\vec #1}
\def\o{\omega}
\def\O{\Omega}
\def\va{\varepsilon}
\def\D{\nabla}
\def\.{\cdot}
\def\x{\times}
\def\l#1{\label{eq:#1}}
\def\r#1{(\ref{eq:#1})}
\def\am{\left(\begin{array}{c}}
\def\amm{\left(\begin{array}{cc}}
\def\a{\end{array}\right)}
\newcommand{\ds}{\displaystyle}

\title{Towards isotropic negative magnetics in the visible range}

\author{C.R. Simovski, S.A. Tretyakov}

\affiliation{Department of Electric and Electronic
Engineering/SMARAD, Helsinki University of Technology, FI-02015,
TKK, Espoo, Finland}

\begin{abstract}
The idea of isotropic resonant magnetism in the visible range of
frequencies known from precedent publications is developed having
in mind achievements of the modern chemistry. Plasmonic colloidal
nanoparticles covering a silica core form a cluster with resonant
and isotropic magnetic response. Two approximate models giving the
qualitative mutual agreement are used to evaluate the magnetic
polarizability of the cluster. It is shown that the electrostatic
interaction of nanocolloids decreases the resonant frequency of an
individual complex magnetic scatterer (nanocluster) compared to
the previously studied variant of a planar circular nanocluster
with same size. This means the reduction of the optical size of
nanoclusters that presumably allows one to avoid strong spatial
dispersion within the frequency range of the negative
permeability.
\end{abstract}

\pacs{78.20.Ci, 42.70.Qs, 42.25.Gy, 73.20.Mf, 78.67.Bf} \maketitle

\section{Introduction}

The interest in artificial resonant magnetism in the optical range
which was recently inspired by the development of metamaterial
science (see e.g. in \cite{Engheta1,Caloz}) is expected to grow
after successful demonstration of the subwavelength optical
imaging obtained in the far zone of objects with the use of
so-called \emph{hyperlenses} \cite{Zhang,Smol}. Hyperlenses
transform evanescent waves into propagating ones and create the
spatially magnified optical images keeping subwavelength details
of the objects.

Known hyperlenses operate with TM-polarized waves. This
polarization is related to the negative permittivity of metals
(which are used by known hyperlenses) in the visible frequency
range. Similar operation with the TE-polarized light would require
materials with negative permeability.

Materials with negative $\mu$ in the visible range do not exist in
nature. Recently, many attempts to engineer such materials were
reported. They include, for example, lattices of paired plasmonic
nanowires, nanoplates, nanocones, fishnets, and plasmonic split
rings \cite{4,5,6,8,9,fishnet,fishnet1,11}. These structures are
all geometrically strongly anisotropic. The strongest magnetic
response corresponds to one specific direction of the propagation
of electromagnetic wave. Notice, that in cited papers the magnetic
response of these structures was reported only for this single
direction. The possibility to use strongly anisotropic artificial
magnetic media in superlenses and hyperlenses has not yet been
studied.

The first work devoted to the possibility of the isotropic optical
magnetism that cannot be achieved in the mentioned anisotropic
structures was, to our knowledge, \cite{Engheta}. The idea of the
present paper is to link this resonant magnetism to the existing
technology and to find the design which ensures the real isotropy
of the resonant permeability, i.e. absence of strong spatial
dispersion. {We use the term strong spatial dispersion for the
case when the dispersion diagram of the lattice is quantitatively
different from the dispersion in any continuous media. Weak
spatial dispersion is a special term applied to designate effects
of artificial magnetism and bianisotropy in composite media
\cite{biaha}.}

In \cite{Engheta} one suggested to prepare optical magnetic media
using plasmonic (colloids of noble metals) nanospheres arranged in
nanorings. In high-frequency magnetic field ${\bf H}_0$
nanocolloids are polarized in the azimuthal direction with respect
to this field. To obtain an isotropic resonant magnetic medium one
should fabricate a dense array of nanorings lying in 3 mutually
orthogonal planes. The dense isotropic package means that the
array of nanorings must be a fcc or a bcc lattice, that allows us
to consider the structure as a photonic crystal. The theoretical
prediction of \cite{Engheta} shows that in this case the resonance
of the effective permeability $\mu_{\rm eff}$ can be strong enough
so that the real part of $\mu_{\rm eff}$ takes negative values.
However, from \cite{Engheta} it has not yet been clear how to
fabricate these structures and their isotropy was not proved. In
fact, an isotropic artificial magnetism is not possible for
photonic crystals (lattices with strong spatial dispersion).

Formally, the resonant magnetic permeability as well as the
permittivity can be attributed to photonic crystals as well (see
e.g. \cite{bred4,13,14}). However, at frequencies where a cubic
lattice becomes a photonic crystal the refraction index and the
wave impedance both become strongly dependent on the propagation
direction \cite{agranovich}. This implies that effective material
parameters $\varepsilon_{\rm eff}$ or/and $\mu_{\rm eff}$ of the
crystal are very different for a wave propagating along one of the
lattice axes and for a wave tilted with respect to this axis. This
evidently means the absence of the isotropy. Of course, the random
mixture of particles is more optically isotropic than the lattice
with the same concentration of particles. However, in the present
case the requirement of the negative magnetic response of the
composite medium obviously implies the very dense packaging of
magnetic scatterers. In the present case we cannot suppress the
spatial dispersion using the randomization of the array.

The comparison of the approximate dispersion diagram for the
homogenized lattice with the exact dispersion diagram was done in
some papers, e.g. in \cite{Li,Belov}. In \cite{Li} the fcc
(face-centered cubic) and bcc (body-centered cubic) lattices of
metal-covered dielectric spheres were studied. It was found that
in fcc and bcc lattices the spatial dispersion effects arise
within the frequency band of the plasmonic resonance. The spatial
dispersion in this region leads to closing the resonance band-gap
with simultaneous strong anisotropy of the lattice refraction
index.  These effects correspond to the frequencies at which the
distance between adjacent particles centers exceeds one quarter of
the wavelength in the matrix.

Unfortunately, in the known literature there are no general
results for the frequency intervals where the effects of strong
spatial dispersion arise in lattices. Of course, it is clear that
the frequency range corresponding to $D/\lambda \ge 0.5$
definitely implies the strong spatial dispersion since
$\lambda=2D$ is the Bragg resonance condition. In lattices of
strongly scattering inclusions the condition of the absence of
strong spatial dispersion becomes more restrictive since the
polarization of inclusions shortens the effective wavelength
$\lambda_{\rm eff}$ compared to $\lambda$.


It is known that the artificial resonant magnetism in arrays of
small complex-shape metal particles at microwaves practically
corresponds to relation $D_p/\lambda_r>0.1$, where $D_p$ is the
typical size of the particle \cite{biaha,biama} and $\lambda_r$ is
the wavelength of the magnetic resonance of a particle in host
medium. Notice, that the ratio $D_p/\lambda\approx 0.1\dots 0.15$
is practically achievable, e.g. for well-known double split-ring
resonators (SRRs). SRRs of this optical size can be considered as
small enough to avoid the strong spatial dispersion in a compact
lattice. Then we can consider the low-frequency stop-band of this
lattice associated with the resonance of SRRs as the effect of the
negative permeability. Really, magneto-electric media are opaque
when their material parameters have different signs\footnote{In
fact, specific effects of strong spatial dispersion like the
excitation of so-called staggered and magneto-inductive modes are
still possible within the resonant stop-band  of the lattice of
optically small SRRs with various design parameters.}.

However, the metamaterial suggested in \cite{Engheta} corresponds
to a quite large optical size of a magnetic scatterer at its
resonance, where the condition $D_p/\lambda_r< 0.15$ is far from
being fulfilled. Really, consider the results for $\mu_{\rm eff}$
illustrated by Fig. 2 (a,b) of \cite{Engheta}. The negative
permeability was theoretically achieved at $630$ or $680$ THz with
nanorings of the following geometry. Six or four metal nanospheres
of radius $a_p=16$ nm are centered at the circle of radius
$R_0=40$ or $R_0=38$ nm. This means that the size $D_p$ of the
effective magnetic particle (i.e. the distance between the edges
of opposite metal nanospheres) is $112$ or $108$ nm. The
wavelength in the host medium with background permittivity
$\va_h=2.2$ at $630$ THz is equal to $\lambda_r=323$ nm. Then the
size of the nanoring from 6 spheres is $D_p\approx 0.35\lambda_r$
i.e. the optical size is rather large. The packaging of these
nanorings in the host medium was assumed in \cite{Engheta} to be
very dense, namely corresponding to their concentration
$N_{NR}=(108)^{-3}$ nm$^{-3}$. This concentration is physically
achievable for a bcc or a fcc lattice of nanorings. However, even
so dense packaging corresponds to the distance between the centers
of adjacent nanorings $D\approx 0.4\lambda_r$. Having in mind
results of \cite{Li} we believe that in this case the strong
spatial dispersion arises and the isotropic magnetism is
impossible.

The clear way to decrease spatial dispersion in a lattice of
magnetic scatterers i.e. to suppress the angular dependence of
$\mu_{\rm eff}$ is to decrease the size of a scatterer keeping its
magnetic polarizability. In this way we will also decrease the
optical distance between the centers of adjacent magnetic
particles. Below we modify the design of magnetic particles
suggested in \cite{Engheta} in this way. We obtain the ratio
$D_p/\lambda_r\approx 0.2\dots 0.25$ which is presumably below the
range of the strong spatial dispersion. Simultaneously, we link
our design to existing nanotechnologies.

\section{Core-shell magnetic clusters}

The design of the optical magnetic scatterer suggested in this
paper is based on the possibility to fabricate arrays of silver or
gold colloidal nanoclusters located on spherical dielectric cores
in a liquid host medium. The fabrication technology uses
self-assembled adhesion of metal nanospheres on the silica core.
The idea of such self-assembling aggregates was probably first
realized in \cite{Morn}. However, this process led to the complete
covering of the silica core by contacting gold nanoparticles. The
process chemically stimulated by some acids which allowed one to
control its speed and to obtain the needed number of nanocolloids
per one core was described in \cite{Mornet}. Later, nanoclusters
comprising few tens of colloidal particles on a silica core of
diameter of the order 100 nm were reported in works \cite{Jiang}.
For obtaining the maximal electromagnetic response for given sizes
(of the core and of the nanocolloids) their number $N_{\rm tot}$
per one nanocluster should be as large as possible. However, it is
difficult in this way to prevent the touching of nanocolloids
which is destructive for their plasmonic resonance that will
degrade due to the tunneling of electrons between nanocolloids.
Practically, minimal separation $d$ between nanocolloids should be
comparable with the radius of colloidal particles $a_p$. Below we
respect the condition $d\approx a_p/2$ that restricts $N_{\rm
tot}$. This can be practically obtained using the technology
described in \cite{new1} and \cite{new2}. In these works the
silica core of diameter 100 nm (the technology allows in principle
to reduce this size to 20-30 nm) was covered by mutually touching
polystyrene nanospheres. Instead of simple polystyrene nanospheres
one can use core-shell particles with plasmonic nanocolloids as
cores. This method will allow us to reliably control the
separation $d$ between colloidal particles which is simply equal
to the double thickness of the polystyrene shell.

Such a nanocluster is shown in Fig.~\ref{fig1} (a) (a general
view) and (b) (a cross section). It should have the isotropic
resonant response to the electromagnetic field of light. At one
frequency the induced electric dipole dominates, and the
polarization of colloidal particles is parallel to the electric
field of light. At another frequency the induced magnetic dipole
dominates, and the polarization of colloidal particles is
azimuthal with respect to the magnetic field. In other words, the
applied magnetic field form the effective nanorings of colloidal
spheres glued to the silica core. Both electric and magnetic
resonant frequencies origin from the plasmonic resonance of an
individual colloidal nanosphere. In this paper we study only the
magnetic resonance of such a nanocluster. Therefore we call it a
magnetic nanocluster (MNC). Bulk arrays of MNC can be created not
only in liquid but also in porous matrices where the separation
between the pores should be small enough to obtain the high
concentration of MNC and a strong magnetic response of the medium.

Below we will see that the frequency of the magnetic resonance of
MNC is strongly reduced compared to the frequency $\O_p$ of the
plasmonic resonance of a single nanocolloid. This effect results
from the strong electrostatic coupling between the nanocolloids in
the MNC. This electrostatic coupling exists also in a single
nanoring studied in \cite{Engheta}, and it leads to a certain
reduction of the magnetic resonance frequency compared to $\O_p$.
However, in the present geometry this reduction is much stronger
since the most important electrostatic interaction is that between
colloids having parallel electro-dipole moments.

\begin{figure}
\begin{center}
\includegraphics[width=125mm]{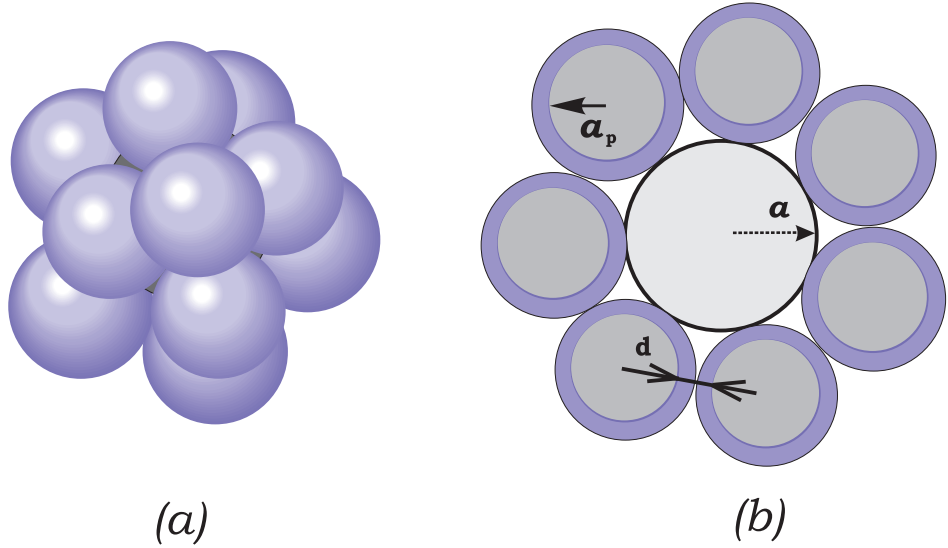}
\caption{A magnetic nanocluster (a bulk isotropic scatterer with
the magnetic resonance): (a) -- a general view of a cluster of
metal nanocolloids with dielectric core, (b) -- the resonant
azimuthal electric polarization $\-P_{\phi}$ is excited in the
discrete "shell" formed by plasmonic nanocolloids when the applied
field is the optical magnetic field directed along $z$.}
\label{fig1}
\end{center}
\end{figure}

The total number $N_{\rm tot}$ of colloidal nanospheres per one
MNC can be controlled by the self-assembling technological
process. Also, it can be adjusted by properly chosen initial
concentrations of nanocolloids and silica spheres in the host
liquid. $N_{\rm tot}$ can be expressed through the core radius
$a$, the colloid radius $a_p$ and the separation $d$ between metal
spheres (see Fig. \ref{fig1}): \e N_{\rm tot}=\left[{4\pi
(a+a_p+d/2)^2\over (2a_p+d)^2}\right]. \l{Ntot}\f Here $[A]$
denotes the integer part of the number $A$. In this formula the
curvature of the portion $\Delta$ of the spherical surface with
radius $R_0=a+a_p+d/2$ which is cut of the sphere of radius $R_0$
(this sphere centers the nanocolloids and their dipole moments are
located on it) by one core-shell particle of radius $a_p+d/2$ is
neglected. In other words, in our calculations the surface
$\Delta$ of the effective sphere $R_0$ per one colloidal particle
is assumed to be a planar square with side $2a_p+d$.

\section{Two models of MNC}

\begin{figure}
\begin{center}
\includegraphics[width=125mm]{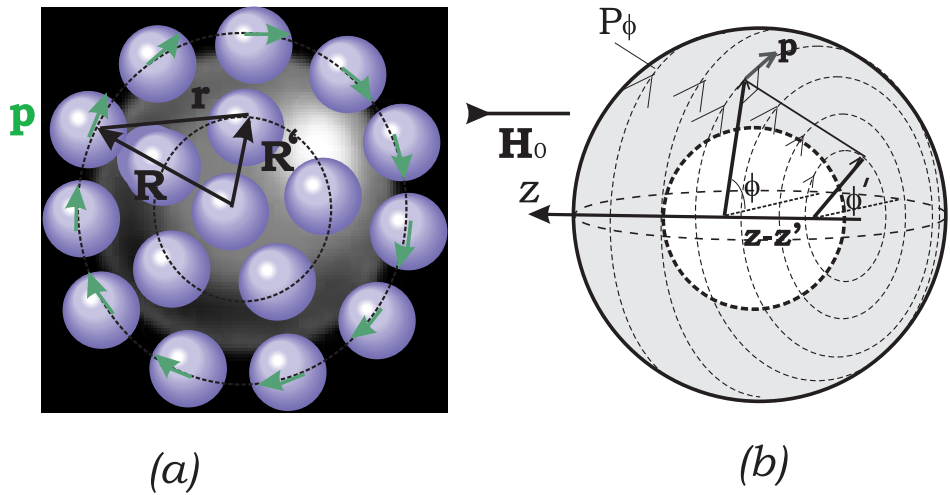}
\caption{Two models of the magnetic polarization of MNC in the
applied magnetic field directed along $z$: (a) -- nanocolloids
form regular rings around the $z-$axis, the colloidal particles
are equidistant in the rings of radiuses $R_j$ (induced dipole
moments are shown by bold arrows) and the rings are equidistant;
(b) -- nanocolloids form around the core (shown as a dashed
circle) an effective continuous shell with azimuthal bulk
polarization $P_{\phi}$. The shell of thickness $2a_p+d$ can be
characterized by resonant isotropic permittivity $\va_L$.}
\label{fig2}
\end{center}
\end{figure}

Two approximate models for obtaining the magnetic polarizability
$a_{mm}$ of a MNC are illustrated by Fig.~\ref{fig2} (a) and (b).
The agreement of their results can be hopefully considered as a
validation of these results. These models describe in two
different ways the electromagnetic interaction of colloidal
particles. In the first model the nanocolloids on the dielectric
core are assumed to form regular rings around the $z-$axis. The
rings of radiuses $R_j$ are distanced by $a_p/2$ from one another.
The distance between the nanoparticles in any ring is also equal
to $d=a_p/2$. The presence of the polystyrene shell of
nanocolloids does not influence to the result, since the
permittivity of the shell is equal to that of the matrix
($\varepsilon_h=2.2$).

In the second model the discrete structure of core-shell plasmonic
nanoparticles on the central silica core is treated as a nanolayer
of effective medium of thickness $\delta=2a_p+d$ covering the
silica core. The azimuthal electro-dipole polarization $P_{\phi}$
and the isotropic permittivity $\va_L$ of this layer can be easily
calculated and we find $a_{mm}$ in the closed form calculating the
magnetic moment of this homogenized shell.

The presence of the central silica core also does not influence to
the result due to the absence of its electric polarization by the
applied magnetic field and smallness of its magnetic moment which
is not resonant. It was checked analytically that the interaction
of this small magnetic moment with electro-dipole moments of
colloidal particles is negligible. The permittivity of the silica
core $\va_s=4>\va_h=2.2$ influences only weakly to the response of
colloidal spheres. The polarizability  $\alpha$ of a single
colloidal particle is assumed to be the same as if nanocolloids
were located in the uniform medium with permittivity $\va_h$: \e
\alpha=\left[\left(4\pi a_p^3\va_0\va_h{\va_m-\va_h\over
\va_m+2\va_h} \right)^{-1}-i{k^3\over 6\pi\va_0\va_h}\right]^{-1}.
\l{ind}\f Here $\va_m$ is the permittivity of the metal which is
taken in the same form as in \cite{Engheta}: \e \va_m=\va_i-{(2\pi
f_p)^2\over \o^2}+i{(2\pi)^2 f_pf_d\over \o^2}. \l{epss}\f For
silver colloids we have the plasma linear frequency $f_p=2175$
THz, the damping frequency $f_d=4.35$ THz and the ultraviolet
permittivity $\va_i=5$ \cite{Engheta}.

As it was explained in \cite{Engheta} the magnetic moment of
nanorings is excited by the magnetic field of the light wave. To
calculate the magnetic polarizability of a MNC we assume that it
is excited by magnetic field polarized along $z$ [see
Fig.~\ref{fig1} (b)] that in the model of the magnetic excitation
assume to be uniform across the circle of radius $R_0$ and varying
with time as $\exp(-i\omega t))$ with amplitude $H_0$. From
Maxwell's equations this applied magnetic field $\-H_{a}$ is
associated with the azimuthal electric field $\-E_{a}$ vanishing
at the center of MNC: \e \-H_{a}=\-z_0H_0,\quad
\-E_{a}=\-\phi_0(i\eta H_0 k R), \l{applied}\f where
$\eta=\sqrt{\mu_0/\va_0\va_h}$ is the wave impedance of the host
medium of permittivity $\va_h$. Unit vectors $\-z_0$ and
$\-\phi_0$ correspond to the cylindrical coordinates $(R,\phi,z)$
shown in Fig.~\ref{fig1} (b). In fact, the applied electric field
given by \r{applied} is associated with slightly non-uniform
magnetic field $\-H_{a}=\-z_0 H_0(1-k^2R^2/2)$, but this small
correction to the applied magnetic field matters nothing for our
model. We calculate the magnetic moment of MNC induced by
$\-H_{a}$ whose amplitude at the center of MNC is $H_0$ through
the electric polarization of nanocolloids by the electric field
associated with $\-H_{a}$.

The $\phi$-oriented dipole moments $p_j$ of colloidal particles of
any $j$-th ring are identical in polar coordinates. The
$z$-directed magnetic moment $m_{j}$ of the j-th ring can be found
using formula (8) of \cite{Engheta}. The total magnetic
polarizability of MNC is then obtained as $a_{mm}=\sum
a_{mm}^{(j)}$, where $a_{mm}^{(j)}=m_j/H_0$.

Notice, that the obvious (not sufficient) condition of the absence
of spatial dispersion in a bulk array of MNC is the independence
of the magnetic polarizability $a_{mm}$ of an individual MNC on
the distribution of the applied magnetic field around the center
of MNC, if the condition of the magnetic excitation is respected:
$\-H_{a}(R=0)=\-z_{0}H_0,\ \-E_{a}(R=0)=0$. This condition can be
satisfied for example by a standing plane wave, i.e. a pair of
plane waves travelling along the $x-$axis:
$$
\-H_{a}=\-z_0{H_0\over 2}(e^{-jkx}+e^{jkx}),\quad
\-E_{a}=\-y_0{\eta H_0\over 2}(e^{-jkx}-e^{jkx}).
$$
The comparison of $a_{mm}$ calculated for this azimuth-dependent
excitation of MNC and that calculated for excitation \r{applied}
would be a useful exercise, and we plan it for the next paper. It
could indicate the frequency bounds of the spatial dispersion
related to the finite size of MNC. However, in the present paper
we restrict the analysis by a more simple excitation case
\r{applied}.

In both models the reduction of the resonance frequency of MNC
compared to the plasmonic resonance $\O_p$ of the individual
colloidal particle in the host medium with $\va_h=2.2$ is
determined by the electromagnetic coupling of colloids. In fact,
this reduction will be even stronger than it is predicted below.
This is because the presence of the silica core makes the
effective host medium permittivity slightly larger than $\va_h$.
We neglect this small effect for simplicity of the model, however,
its role is positive.

\subsection{The model of regular rings}

Consider a dipole $\-p=p(j)\-\phi_0$ shown in Fig.~\ref{fig1} (b)
that belongs to the j-th effective ring of MNC. Assume that
$N_{\rm tot}\gg 1$ (practically the model is applicable when
$N_{\rm tot}>15\dots 20$). Then it is easy to show using auxiliary
spherical coordinates and formula \r{Ntot} for $N_{\rm tot}$ that
the radius of the j-th ring is approximately equal to $R_j\approx
R_0\sin({2j\sqrt{\pi/ N_{\rm tot}}})$. The $z-$coordinate of the
j-th ring is $z(j)\approx R_0\cos({2j\sqrt{\pi/N_{\rm tot}}})$.
The number of dipoles $N_j$ in the ring is equal $N_j\approx 2\pi
R_j/(2a_p+d)$. The number of such rings in MNC is equal
$N_r\approx \sqrt{\pi N_{\rm tot}}/2$. The cylindrical coordinates
of the dipole $\-p$ in the j-th ring are $z=z_j$, $R=R_j$,
$\phi=2\pi q/N_j$ where $q=0\dots (N_j-1)$ is the integer number
determining the position of the dipole within the ring.

The field produced by the dipole $\-p$ at the point with
coordinates $(z',R',\phi')$ can be found from the standard
formula: \e \-E(\-R,\-R')={1\over
4\pi\va_0\va_h}\left[k^2(\-r\x\-p)\x \-r{e^{-jkr}\over
r^3}+(3\-r\-p\cdot\-r-\-pr^2)\left({e^{ikr}\over
r^5}-{ike^{ikr}\over r^4}\right) \right]. \l{nnn}\f Here
$r=\sqrt{R^2+R^{'2}-2RR'\cos(\phi-\phi')+(z-z')^2}$ is the
distance between radiating $p(j)$ and receiving $p(n)$ dipoles of
MNC. Due to the azimuthal symmetry of the problem only the
$\phi-$th component of the field produced by all rings of dipoles
$p(j)$ is nonzero at the center of the receiving dipole $p(n)$,
i.e. at point $(z',R',\phi')$. In other words the local electric
field acting on the dipole $p(n)$ is orthogonal to the vector
$\-R'$ shown in Fig. \ref{fig1} (b). Respectively, only the
azimuthal component of the vector $\-E(\-R,\-R')$,  should be
taken into account.

From \r{nnn} the scalar interaction coefficient of dipoles $p(j)$
and $p(n)$ can be easily derived. It is defined as the $\phi$-th
component of the field produced by the unit azimuth-oriented
dipole located at point $(z=z_j,R=R_j,\phi=2\pi q/N_j)$ and
calculated at point $(z'=z_n,R'=R_n,\phi'=2\pi s/N_n)$:
$$Q_{qs}^{nj}={e^{ikr}\over
4\pi\va_0\va_hr^5}\left[(kr)^2(R^{'2}+(z-z')R'-RR'+\cos(\phi-\phi')(R^2+(z-z')R)-RR')
\right. $$ \e \left.
-3RR'\sin^2(\phi-\phi')(1-ikr)-\cos(\phi-\phi')(1-ikr)r^2
 \right]. \l{Q}\f
The dipole moment of the receiving dipole is equal to $ p=\alpha
E^{\rm loc}$. Here the local field is the sum of the external
electric field \r{applied} $E_a(R')=(i H_0\eta k R'/ 2)$ and all
dipole fields:
$$E^{\rm loc}=E_a(R')+p(j)\sum\limits_{s,j} Q_{qs}^{nj}.$$
This way we obtain the system of equations for dipole moments of
colloidal nanospheres of any ring: \e {1\over
\alpha}p(n)={iH_0\o\mu_0 R(n)\over 2}+
p(n)\sum\limits_{q=1}^{N_n-1}Q_{0q}^{nn}+\sum\limits_{j\ne
n}^{N_r}p(j)\sum\limits_{q=0}^{N_j-1}Q_{0q}^{nj}. \l{system}\f
Here the term with $Q_{0q}^{nn}$ describing the interaction of the
dipole $p(n)$ located at ($z'=z_n, R'=R_n, \phi'=0$) with other
dipoles of the same n-th ring is shared out. The expressions for
coefficients $Q_{0q}^{nn}$ entering this term and given by \r{Q}
with $z=z'$ are described by \r{Q}. There is no coefficient
$Q_{00}^{nn}$ in \r{system} since the summation starts from $q=1$
(see also in \cite{Engheta}).

Solving the system \r{system} we find all dipole moments $p(n)$.
The magnetic moment of $j$-th ring is calculated as in
\cite{Engheta}: \e m_j={-i\o p(j)N_jR_j\over 2}. \l{M}\f Then we
obtain the magnetic polarizability of the MNC as the sum of $m_j$
letting $H_0=1$:
$$
a_{mm}={{-i\o\over 2}}\sum\limits_{j=1}^{N_r}p(j)N_jR_j.
$$
The relative effective permeability of the composite medium is
given by \cite{Engheta}: \e \mu_{\rm eff}=1+{1\over
N_{MNC}^{-1}a_{mm}^{-1}-{1\over 3}}. \l{mu}\f Here $N_{MNC}$ is
the volume concentration of MNC that can be expressed though the
effective volume per one magnetic scatterer
$N_{MNC}=1/V_{0}=1/D^3$. In simple cubic lattices $D\ge D_p=
2(a+2a_p)$, and $D$ is the unit cell size. However a more compact
arrangement of MNC is also possible when $D<D_p$ and
$N_{MNC}>1/D_p^3$.

The inverse polarizability of a nanocolloid in \r{system}
corresponds to formula \r{ind} and contains the term
$(-i{k^3/6\pi\va_0\va_h})$ that describes the radiation damping.
The radiation damping of the magnetic dipole with magnetic
polarizability $a_{mm}$ should be described by the term
$(-i{k^3/6\pi})$ \cite{Engheta}. It is known that the radiation
damping is cancelled out in regular 3D arrays and for regular
lattices we should have used instead of \r{mu} the relation
\cite{Engheta}: \e \mu_{\rm eff}=1+{1\over
N_{MNC}^{-1}\left(a_{mm}^{-1}+i{k^3\over 6\pi}\right)-{1\over 3}}.
\l{mu1}\f However, the dissipative losses due to the plasmonic
resonances of metal nanospheres strongly dominate over the
contribution of radiation losses into the imaginary part of the
permeability, i.e. the difference in results of \r{mu} and of
\r{mu1} is negligible.

\subsection{The model of an effective shell}

Replacing the discrete plasmonic structure by the continuous shell
we introduce the bulk polarization $P_{\phi}$ that can be
expressed through the averaged $\phi-$polarized electric field
$E_{\rm av}$ in this shell and its unknown permittivity $\va_L$:
\e P_{\phi}=\va_0(\va_L-\va_h)E_{\rm av} \l{eee}\f The field
$E_{\rm av}$ is related with the local field acting on any
colloidal nanosphere by the Clausius-Mossotti relation: \e E_{\rm
av}=E_{\rm loc}-{p\over 3V_1\va_0\va_h}. \l{CM}\f Here
$V_1=2a_p(2a_p+d)^2$ is the volume per one colloidal nanosphere
and $p=P_{\phi}V_1$ is the dipole moment of the reference
nanosphere. Using the formula $p=\alpha E_{\rm loc}$ together with
\r{eee} and \r{CM} we come to
 the Lorentz-Lorenz formula: \e
\va_L=\va_h\left(1+{3\over {3\va_0\va_h V_1\over\alpha
}-1}\right). \l{epsL}\f

The definition of the magnetic moment of any volume $V$ comprising
polarization currents $\-j$ reads as:
$$
m={1\over 2}\int\limits_V \-j\x \-r dV.
$$
It can be rewritten for the MNC in terms of the bulk polarization
$P_{\phi}$: \e m= {-i\o \over 2}\int\limits_{V_L}P_{\phi}R\, dV.
 \l{ppp}\f
Here $V_L=4\pi R_0^2(2a_p)=8\pi a_p(a+a_p)^2$ is the volume of the
spherical layer with central radius $R_0$ and the thickness
$2a_p$. The integration of the bulk polarization across this layer
can be replaced by simple product $P_{\phi}(2a_p)$. The polar
radius $R$ that enters \r{ppp} can be expressed in spherical
coordinates as $R(\theta)=R_0\sin\theta$. After substitution of
\r{eee} into \r{ppp} we obtain: \e m= {-i\o\va_0
a_p(\va_L-\va_h)R_0^2 }\int\limits_{0}^{2\pi}\, d\phi
\int\limits_{0}^{\pi}\, d\theta \sin\theta E_{\rm
av}(\theta)R(\theta).
 \l{pp}\f
The averaged electric field of azimuthal polarization at the
circle of radius $R(\theta)$ is related to the magnetic field
$H_0$ at the center of the MNC as $E_{\rm
av}(\theta)=i\o\mu_0R(\theta)H_0/2=(i H_0\eta k R_0 \sin\theta/
2)$, and after this substitution into \r{pp} and trivial
integration we come (letting $H_0=1$) to the following formula: \e
a_{mm}=m= {4\pi\over 3}(k_0R_0)^2V_0(\va_L-\va_h), \l{p}\f where
it is denoted $V_0=R_0^2a_p$ and
$k_0=k/\sqrt{\va_h}=\omega\sqrt{\va_0\mu_0}$ is the free space
wave number.

Substituting \r{epsL} into \r{p} we obtain the final closed-form
formula for the magnetic polarizability of individual MNC: \e
a_{mm}= {4\pi}(k_0R_0)^2V_0{\va_h\over {{3\va_0\va_h
V_1\over\alpha }-1}}. \l{mmm}\f The permeability can be then found
using \r{mu} or \r{mu1}.

\section{Results and discussion}

First, the results obtained in \cite{Engheta} were reproduced
using the first model (the second model implies very large
$N_{tot}$ and cannot be applied to the case of a single ring in
MNC). Namely, the system \r{system} was solved for $N_r=1$, when
it was taken also $N_{tot}=4$, $R_0=38$ nm and $a_p=16$ nm. This
geometry corresponds to the total size of the magnetic cluster
$D_p=108$ nm and to the spherical core radius (recall, that in the
present theory the difference of the core permittivity from that
of the host medium is not taken into account) $a=R_0-a_p=22$ nm.
The concentration $N_{MNC}$ of effective magnetic scatterers was
assumed in \cite{Engheta} to be equal $N_{MNC}=95^{-3}$ nm$^{-3}$
(almost touching MNC in a fcc or a bcc lattice), the metal of
colloidal particles was silver with permittivity \r{epss}.

\begin{figure}
\begin{center}
\subfigure[][]{\label{Y1}\includegraphics[width=7cm]{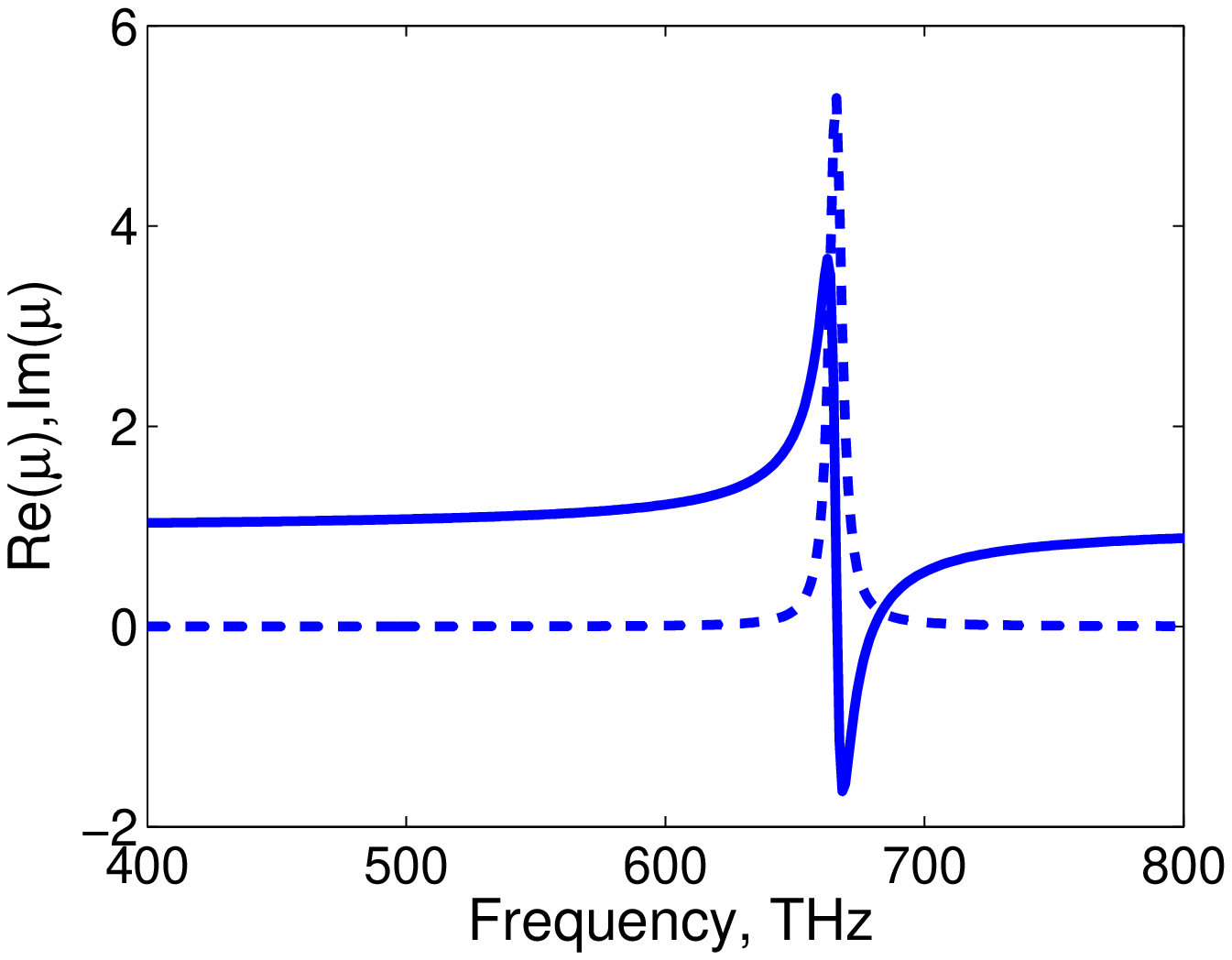}}
\subfigure[][]{\label{Y2}\includegraphics[width=7cm]{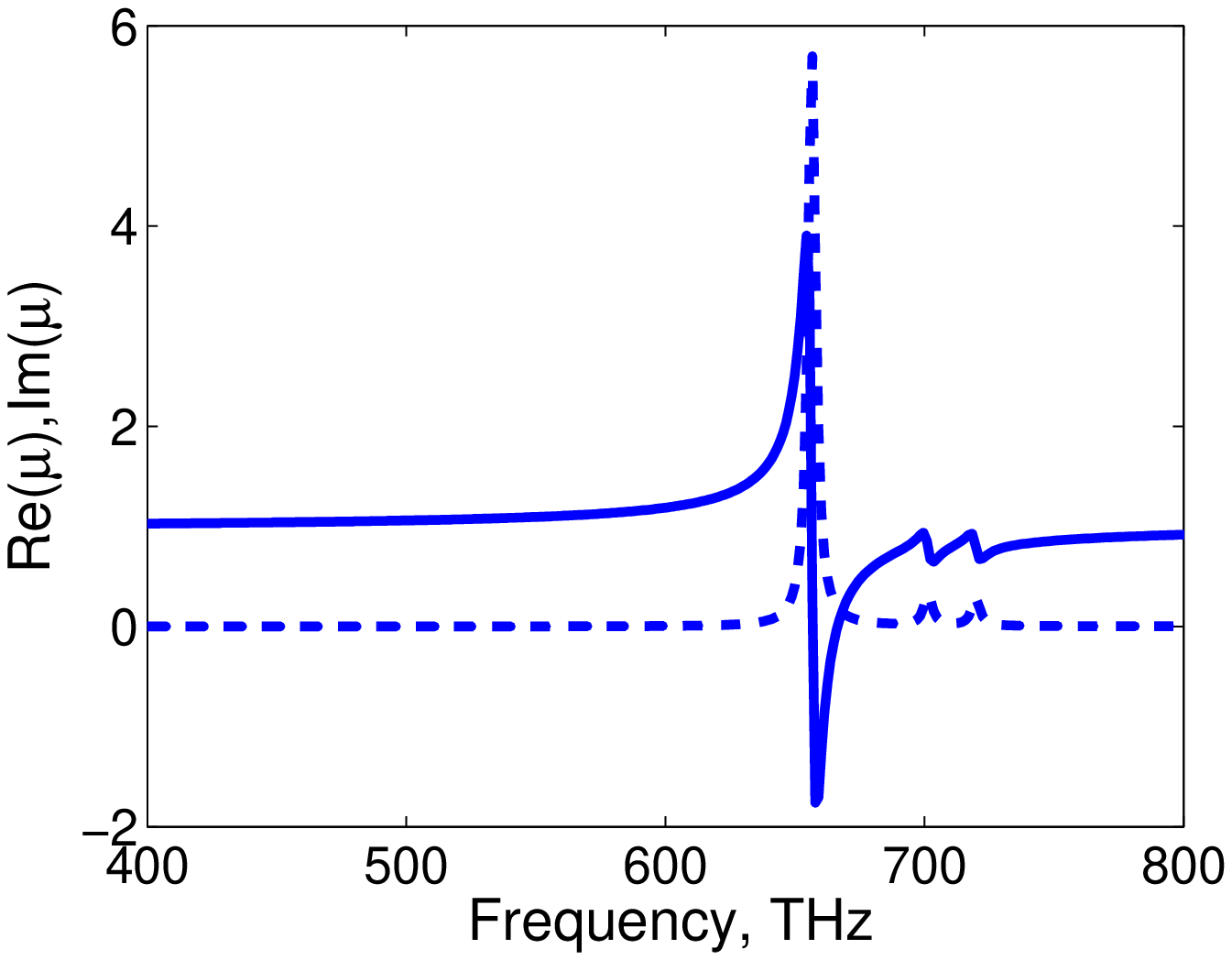}}
\caption{Effective permeability of the array of MNC with sizes
$D_p=108$ nm, $a=22$ nm, $a_p=16$ nm ($R_0=38$ nm) hosted in the
matrix with $\va_h=2.2$. \subref{Y1}: One ring of nanocolloids in
every MNC, the concentration of MNC is $N_{MNC}=95^{-3}$
nm$^{-3}$. \subref{Y2}: Three nanorings in every MNC, other
parameters are the same. Real and imaginary parts of the
permeability are shown by solid and dashed lines, respectively.}
\label{fig3}
\end{center}
\end{figure}

The result for the effective permeability is presented in
Fig.~\ref{fig3} (a). It reproduces  Fig.~2 (b) of \cite{Engheta}
with high accuracy. This agreement can be considered as a check.

\begin{figure}
\begin{center}
\subfigure[][]{\label{Y11}\includegraphics[width=7cm]{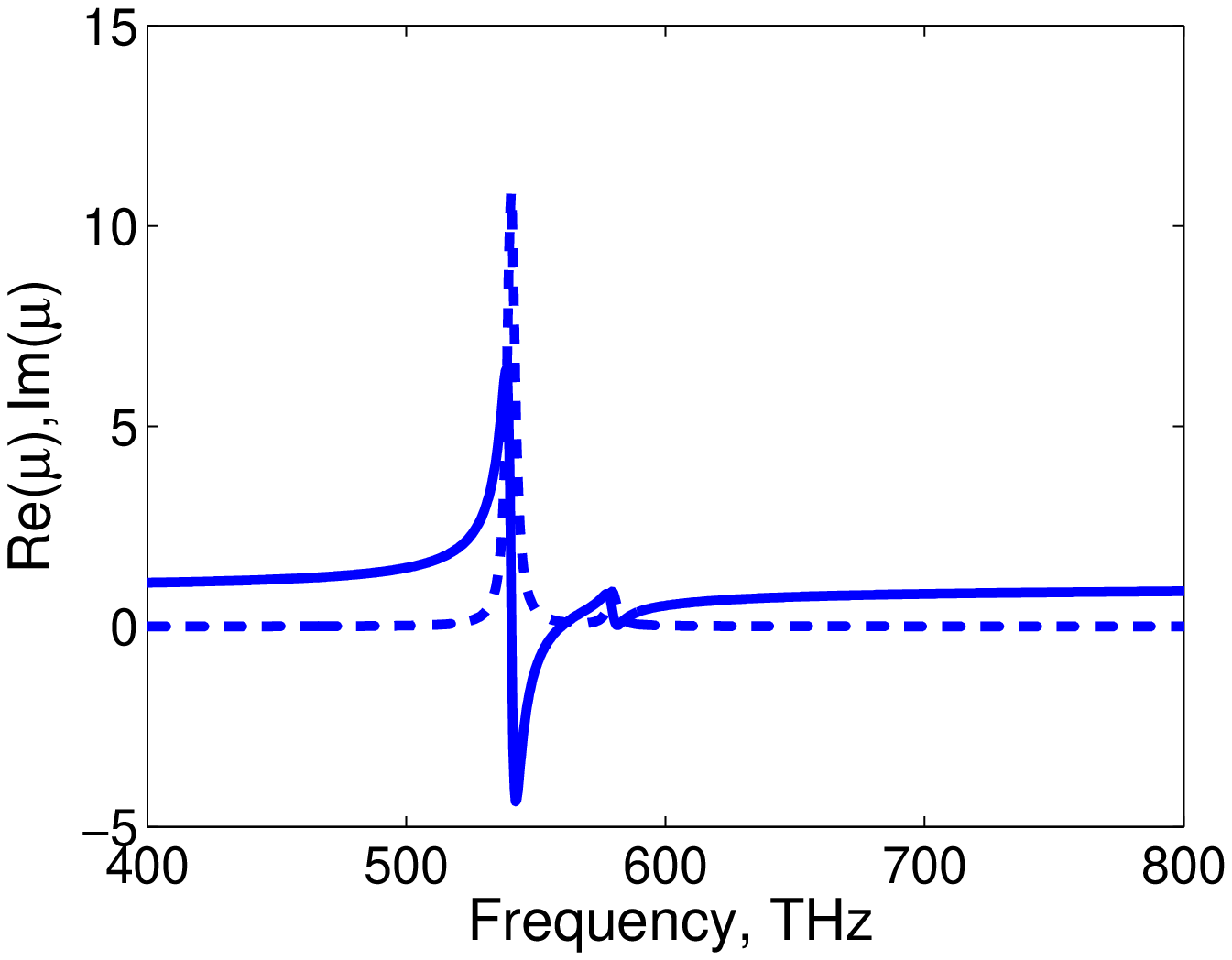}}
\subfigure[][]{\label{Y22}\includegraphics[width=7cm]{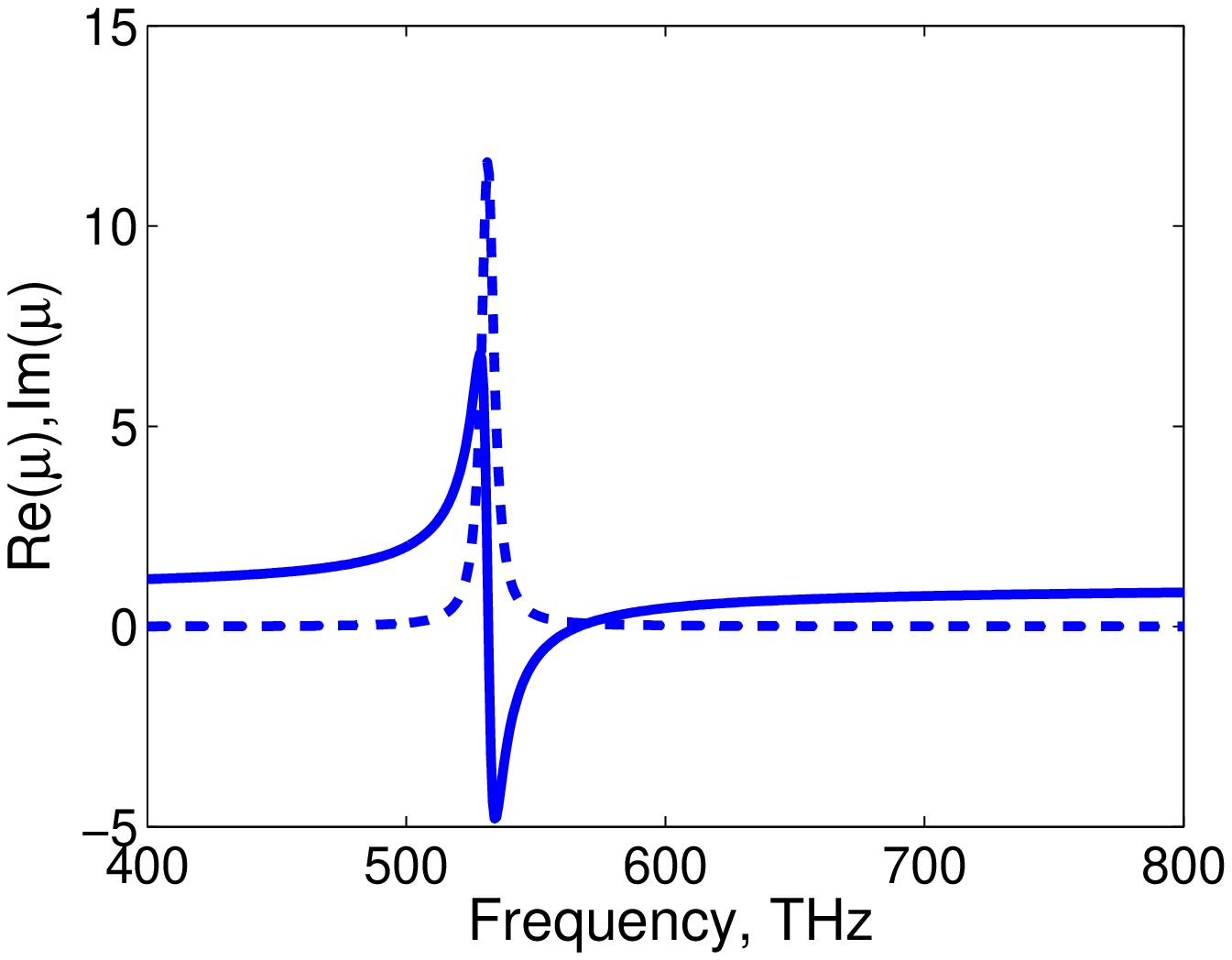}}
\caption{Effective permeability of the array of MNC with sizes
$D_p=108$ nm, $a=22$ nm, $a_p=13$ nm, $d=6$ nm hosted in the
matrix with $\va_h=2.2$, the concentration of MNC is
$N_{MNC}=95^{-3}$ nm$^{-3}$. \subref{Y11}: First model of the
interaction of nanocolloids. \subref{Y22}: Second model of the
interaction of nanocolloids.} \label{fig4}
\end{center}
\end{figure}

If we keep the same separation of colloidal particles $d=10$ nm as
in the previous example the maximal possible number of colloidal
particles for the spherical core of radius that corresponds to
this example is equal to $N_{\rm tot}=16$. It corresponds to
$N_r=3$. Keeping the same concentration $N_{MNC}=95^{-3}$
nm$^{-3}$ of same MNC we obtained with this geometry the result
shown in Fig.~\ref{fig3} (b). This result was also obtained using
the first model (for $N_{\rm tot}=16$ the second model is
inadequate). The reduction of the magnetic resonance frequency due
to the presence of two additional nanorings compared to
Fig.~\ref{fig3} (a) is not dramatic though visible.

\begin{figure}
\begin{center}
\subfigure[][]{\label{Y11}\includegraphics[width=7cm]{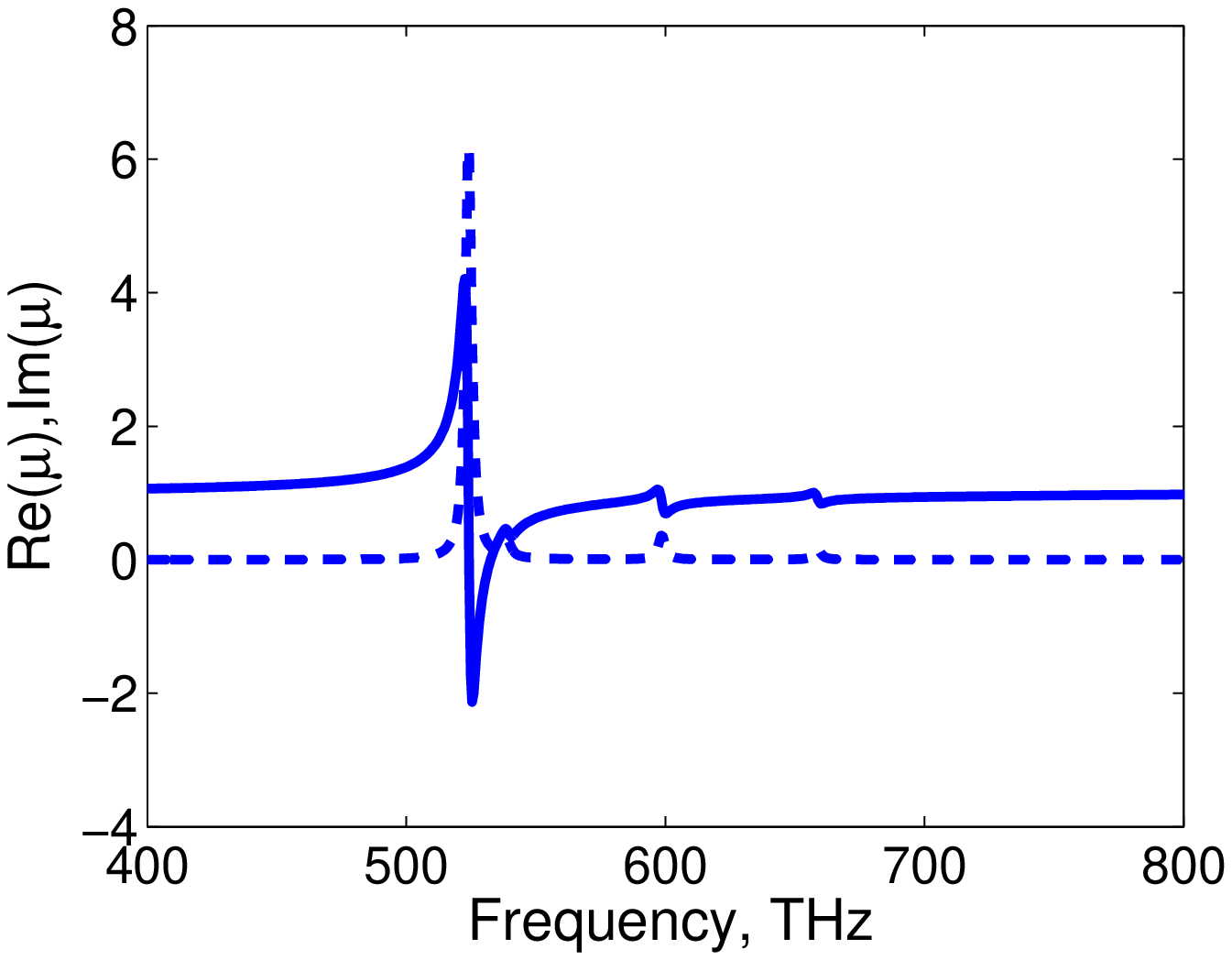}}
\subfigure[][]{\label{Y22}\includegraphics[width=7cm]{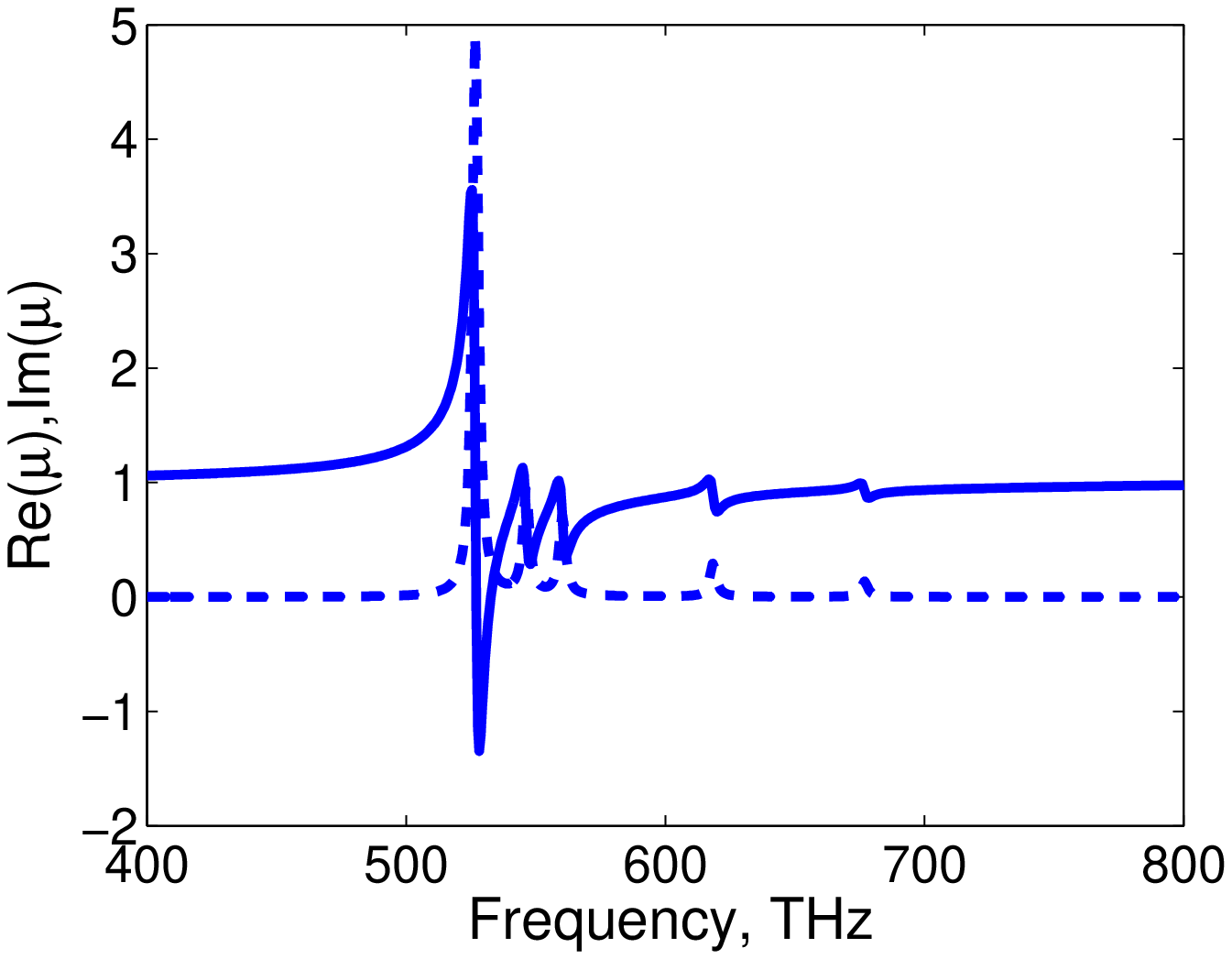}}
\caption{Effective permeability of the array of MNC with sizes
$D_p=108$ nm, $a=22$ nm hosted in the matrix with $\va_h=2.2$, the
concentration of MNC is $N_{MNC}=110$ nm$^{-3}$. \subref{Y11}:
$a_p=13$ nm, $d=6$ nm. \subref{Y22}: $a_p=12.5$ nm, $d=6$ nm.}
\label{fig5}
\end{center}
\end{figure}

More promising results were obtained keeping the same total size
of MNC $D_p=108$ nm for $a_p=13$ nm, $d=6$ nm, that implies $a=22$
nm. This geometry corresponds to one effective ring with $9$
colloids, two rings with $5$ colloids and two rings with $3$
colloids. In this case the number of colloidal particles is large
enough and the qualitative agreement of two models was obtained.
The concentration of MNC in Fig.~\ref{fig4} is the same as in the
previous example: $N_{MNC}=95^{-3}$ nm$^{-3}$. The two models show
approximately the same resonant frequency and the same magnitude
of the resonance. The second model ignores, of course, the higher
magnetic resonances that should arise in interacting nanorings and
are revealed if we use the fist model. However, these higher
resonances are weak and have no practical meaning.

The reduction of the resonant frequency in Fig.~\ref{fig4}
compared to Fig.~\ref{fig3} means that the resonant size of the
MNC reduces from $0.35 \lambda_r$ as in the previous example to
$0.25 \lambda$ in the present one. Here $\lambda_r$ is the
resonant wavelength in the host medium.

High magnitude of the Lorentz resonance in Fig.~\ref{fig4} means
that we can also reduce the concentration of MNC making the array
of MNC easier for manufacturing. The results depicted in
Fig.~\ref{fig5} (a) and (b) are obtained using the first model for
$N_{MNC}=110^{-3}$ nm$^{-3}$ instead of $N_{MNC}=95^{-3}$
nm$^{-3}$ as in the previous case. The difference between two
plots in Fig.~\ref{fig5} (a) and (b) is determined by a 1 nm
difference in the radiuses of plasmonic nanospheres, and
demonstrates how sensible is the magnetic response of MNC to the
deviation of parameters.



\section{Conclusions}

In the present paper we developed the idea of the resonant optical
magnetism in its isotropic variant modifying the known design of
optical magnetic scatterers suggested in \cite{Engheta}. The array
of isotropic magnetic nanoclusters proposed in the present paper
can be obtained in a liquid or porous matrix using the existing
nanotechnologies. The metal nanocolloids are located in the
core-shell particles attached to the silica core. Their
electrostatic mutual coupling helps to reduce the resonant
frequency for the given size of a nanocluster compared to the
frequency of the plasmonic resonance of a single nanocolloid. This
reduction is the essential result of the present paper since it
allows one to decrease the spatial dispersion of the dense array
(lattice) of nanoclusters. In other words, the suggested geometry
is helpful for the isotropy of the effective resonant permeability
of the array. The further optimization of the structure, further
validation of the suggested model by full-wave numerical
simulations are planned in the next paper.




\end{document}